\documentclass[12pt]{article}
\usepackage{amssymb,amsfonts,amsmath}
\usepackage{citesort}
\usepackage[T2A]{fontenc}
\usepackage{graphicx}
\usepackage{braket}
\usepackage{wrapfig}

\newcommand{\NaxCoO}{Na${}_x$CoO${}_2$ }

\begin{document}

\title{Ab-initio studying of spin states of sodium cobaltate Na${}_{2/3}$CoO${}_2$}
\author{Yu. V. Lysogorskiy${}^{1,2}$\thanks{Email: yura.lysogorskii@gmail.com}, \and O.V. Nedopekin${}^2$, \and S. Krivenko${}^2$ \and D.A. Tayurskii ${}^2$, \and B. Minisini${}^1$}
\date{ ${}^{1)}$ Institut Superieur des Materiaux et Mecaniques Avances du Mans, 72000, 44 Av. Bartholdi, Le Mans, France\\
${}^{2)}$ Institute of Physics, Kazan Federal University, 420008, 18 Kremlevskaya st., Kazan, Russia}

\maketitle

\begin{abstract}
Puzzling properties of sodium cobaltates Na${}_x$CoO${}_2$ become an issue of numerous recent theoretical and experimental investigations because of very high thermopower and intricate phase diagram of this material. Experiments indicated Na${}_{2/3}$CoO${}_2$ to have a kagome-like cobalt sublattice, which implies additional complexity of this strongly correlated system. In the present work we have employed the ab-initio GGA+U modeling method to investigate spin states of Na${}_{2/3}$CoO${}_2$, using an exact crystal cell established in the experiment. Stability of sodium imprint for concentration $x=2/3$ proposed in the publications has been obtained. An AFM-A type magnetic ordering has been recognized as a ground state. Moreover, a competition between low spin state (LS) and intermediate spin state (IS) has been established to depend on on-site Hubbard $U$ correction parameter. LS state has been identified as a ground state of Na${}_{2/3}$CoO${}_2$.
\end{abstract}

\maketitle

\section{Introduction}

The intricate ground-state properties of the sodium cobaltates Na${}_x$CoO${}_2$ are a subject of many recent theoretical and experimental investigations. This material holds much promise for thermoelectronics due to its large thermopower~\cite{Fujita2001} together with the relatively low resistivity~\cite{Terasaki1997}. Magnetic fields strongly affect the thermopower in this material, which suggests that electron-electron strong correlations and spin degrees of freedom are important. The discovery of superconductivity with T${}_c$ about 5 K in Na${}_{0.33}$CoO${}_2$ $\cdot$ 1.3 H${}_2$O \cite{takada2003superconductivity} revived the interest to the physics of the lamellar sodium cobaltates providing the first known example of a non-cuprate oxide superconductors.

Such properties as the superconductivity~\cite{takada2003superconductivity}, high thermopower coefficient~\cite{Terasaki1997}, uncertainty of the charge state of the cobalt ions in triangular crystal lattice and strong influence of the sodium ions arrangement onto electronic properties of CoO${}_2$ planes~\cite{Marianetti2007a} make the investigation of sodium cobaltates very important for the modern physics of condensed matter.

Structurally, \NaxCoO consists of quasi two dimensional CoO${}_2$ layers stacked along the crystallographic $c$-axis, separated by Na layers. Forming the triangular lattice, the planar  Co ions are surrounded by edge-sharing O${}_6$ octahedrally, which are trigonally distorted along $c$.

The global $T$-$x$ phase diagram of sodium cobaltate \NaxCoO~\cite{Lang2008} has two regimes - with ferromagnetic and antiferromagnetic correlations. In addition, there are narrow insulating state at $x =1/2$, which separates two distinct metallic regions, as well as superconducting state at $x\approx0.3$. A phase transition between two regimes of magnetic correlations appears at $x^* \approx 0.65$ at low temperatures. At $x<x^*$ the spin dynamics display predominantly antiferromagnetic correlations at low $T$, contrary to the ferromagnetic ones at $x>x^*$.  An origin of this magnetic transition remains a challenging problem, which motivates the present study.

From the experiments it is known that phase above $x^* \approx 0.65$ corresponds to long range AFM order of A-type, i.e., spins are ordered ferromagnetically in CoO${}_2$ plane, whereas exchange integral between planes is antiferromagnetic. It is supposed, that spin and charge ordering of Co${}^{3+}$/Co${}^{4+}$ at $x=2/3$ are due to the simultaneous Na${}^{+}$ ordering and magnetic transition in CoO${}_2$ plane in metallic state. But the mechanisms of a such complex transition still remains the open problem.

Minor deviations of the local electrostatic potential on cobalt sites, such as those induced by local Na${}^+$ order or disorder, may therefore significantly influence the electronic state \cite{Marianetti2007a}. Thus inplane sodium ordering is in a great importance. In Ref.~\cite{Meng2008a,Hinuma2008,Meng2005} sodium ordering patterns for different sodium concentration ($x=0.5$, $0.6$, $2/3$, $0.75$) were proposed. To predict the ground state, authors have used Monte Carlo simulations with GGA cluster expansion method. But in bulk samples not only inplane ordering, but also the relative position of sodium planes is very important. In the next section influence of the sodium ordering will be discussed in more details.

\subsection{Electronic structure}
 In the cobaltate the electronic $3d$ level of the Co ion is splitted by a cubic crystal field of oxygen octahedron into the lower $t_{2g}$ triplet and higher $e_g$ doublet orbital state with a crystal gap about 2 eV~\cite{Singh2000}. A trigonal distortion of O${}_6$ octahedra further splits the  $t_{2g}$ triplet into the doublet $e^\prime_{g}$ and singlet $a_{1g}$ state. In a Co${}^{3+}$ valence state with $d^6$ electronic configuration the six lower energy levels ($t_{2g}$) are filled, with a total spin $S = 0$, while Co${}^{4+}$ should only retain one hole in the $t_{2g}$ multiplet, with $S = 1/2$. Experimentally, differentiation between the valencies of cobalt ions can be induced by the sodium ordering~\cite{Alloul2009a}. It was suggested by Alloul \textit{et al}~\cite{Alloul2009a} that such spatial arrangement of Co${}^{3+}$/Co${}^{4+}$ valence state is induced by a specific periodic distribution of Na${}^{+}$ ions, because nearby sodium cation could attract the electron onto the cobalt and fulfill the lower $t_{2g}$ triplet.
However, the microscopic origin of such ordering of the Co valence state and its relation with the simultaneous magnetic transition was not still established.

 Hereafter we will consider different spin configurations of cobalt ions Co${}^{3+}$ and Co${}^{4+}$, notations can be found in the Table~\ref{table:SpinConfig}. There ZS stands for zero spin, HS for high spin, whereas LS stands for low spin and IS for intermediate spin states.

In the present work we investigate an appearance of the unique state at the triangular Co lattice in the cobaltates at $x=2/3$: the metal with the robust ferromagnetic spin and charge (valence) arrangement. To this end we study a role of the coexisting realistic mechanisms: (i) the d-electronic interaction with a  non-uniform electric potential induced by the superlattice of the Na${}^{+}$ ions in Na${}_{2/3}$CoO${}_2$ and (ii) the strong Coulomb repulsion of the electrons in the 3d shells. There are several methods to deal with it - Hubbard model, DMFT and other. We treat the electronic correlations within the Hubbard model and employ an ab-initio calculations within the GGA+U (LSDA+U) approximation to obtain the states of the system.
Ab initio calculations with Hubbard model for strong correlation (LSDA+U) of simplified two dimensional model lead to the absence of small pockets at the Fermi surface~\cite{Zhou2005,Zhang2004}. But the simplified models used in these works  -- for example, a single CoO${}_2$ layer, doped with electrons - do not demonstrate some peculiarities of the electronic ground state, for example, a wide band gap for $x=1/2$. Thus, using a realistic crystal cell as well as taking into account strong correlations of d-electrons is in a great importance.

\begin{table}[h]
  \caption{\label{table:SpinConfig}Spin configurations for cobalt ions Co${}^{3+}$ and Co${}^{4+}$}
\begin{tabular}{@{}lll}
  \hline
   Co${}^{3+}$ & ZS & HS  \\
  \hline
   configuration & $t_{2g}^6$ $e_g^0$ & $t_{2g}^5$ $e_g^1$ \\
  spin & S=0 & S=1 \\
  \hline
   Co${}^{4+}$ & LS & IS \\
  \hline
   configuration & $t_{2g}^5$ $e_g^0$ & $t_{2g}^4$ $e_g^1$ \\
  spin & S=1/2 & S=3/2 \\
  \hline
\end{tabular}
\end{table}

\section{Ab initio modeling methods}
First-principles calculations were performed by means of the spin-polarized DFT+U method with Hubbard correction. Core electron states were represented by the projector augmented-wave method~\cite{Blochl94} as implemented in the Vienna Ab Initio Simulation Package (VASP 5.2.2)~\cite{Kresse96}, which is embedded into MedeA${}^{\circledR}$ interface. The Perdew - Burke - Ernzerhof exchange correlation revised for solids (PBEsol, see Ref. \cite{PBEsol}) and a plane-wave representation for the wave function with a cutoff of 500 eV were used. Both internal coordinates and unit-cell lattice parameters were fully relaxed. The Brillouin zones were sampled with a $5 \times 5 \times 5$ mesh including the gamma point. The density of the mesh for all calculations is approximately one point per 0.004 A${}^{-3}$. SCF convergence criteria was set to $10^{-5}$ eV. To deal with strong correlations  the simplified (rotationally invariant) approach to the LSDA+U, introduced by Dudarev \textit{et al.}~\cite{Dudarev98} was used. The Hubbard $U_\mathrm{eff}=U-J$ value in the Hamiltonian is taken to be in general 5 eV for Co (but for certain cases we provide an analysis for range of  $U$). This value for $U$ is between the values of $U=4.91$ eV for Co${}^{3+}$ and $U=5.37$ eV for Co${}^{4+}$ obtained with first-principles perturbation theory in Li${}_x$CoO${}_2$~\cite{Zhou2004}.   While GGA+U (or LDA+U) is most often used to open up a Hubbard gap in the electronic structure, it was previously shown that in related systems it also has a significant and meaningful effect on the energy. As it removes the self-interaction on the $d$ orbitals, it leads to a strong charge localization and  significant changes in the ground-state structures~\cite{Zhou.PRB.69.201101}.

During our calculations  we have tested two PAW potential for sodium: with seven valence electrons ($2 p^6 3 s^1$) and with one valence electron ($3 s^1$). It was found that, in the first case, all p-electrons are localized on the sodium cation and do not change spin and charge distribution for cobalt ions. Thus the usage of sodium potential with only one valence electron is sufficient.

\section{Calculation model}

\begin{figure}[h]
  \centering
  \includegraphics[height=200px]{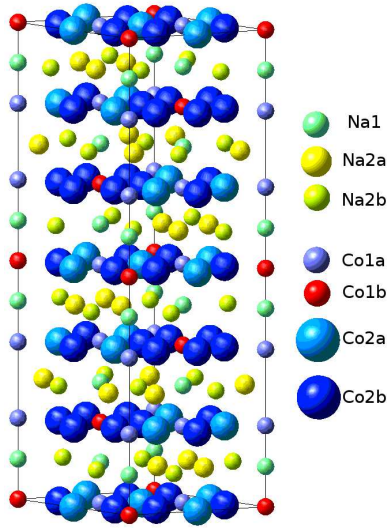}\\
  \caption{Crystallographic cell of Na${}_{2/3}$CoO${}_2$, deduced from NMR/NQR experiments. See~\cite{Alloul2009a} for more details.}\label{fig:Na23CoO2Cell}
\end{figure}

By means of NMR/NQR techniques, which are very useful for dealing with local properties of ions, several unique crystallographic positions of sodium and cobalt ions was identified and minimal crystallographic cell with symmetry group $R-3c$ was built based on this information (Figure~\ref{fig:Na23CoO2Cell})~\cite{Alloul2009a}.

As it can be seen from Figure~\ref{fig:Na23CoO2Cell} the two general crystallographic positions of cobalt ions can be distinguished. First position Co1 (which has two types - Co1a and Co1b) has the minimal distance to the sodium ion. Atom in Co1 positions form a 2D charge ordered state -  a honeycomb network of Co${}^{3+}$ ions with $t_{2g}^6 e_g^0$ electronic configuration. Cobalt ions in another crystallographic positions Co2 (Co2a and Co2b) are forming a kagome lattice of atoms with average $3.44+$ valence state. It means that the electrons are not fully attached to its own ions, but they are hopping  along kagome lattice. Calculation shows that in $a_{1g}$ and $e^\prime_{g}$ orbitals the cobalt electronic states are hybridized with the oxygen one.

\begin{figure}[h]
\begin{minipage}[h]{0.49\linewidth}
 \textit{(a)}
  \center{\includegraphics[height=100px]{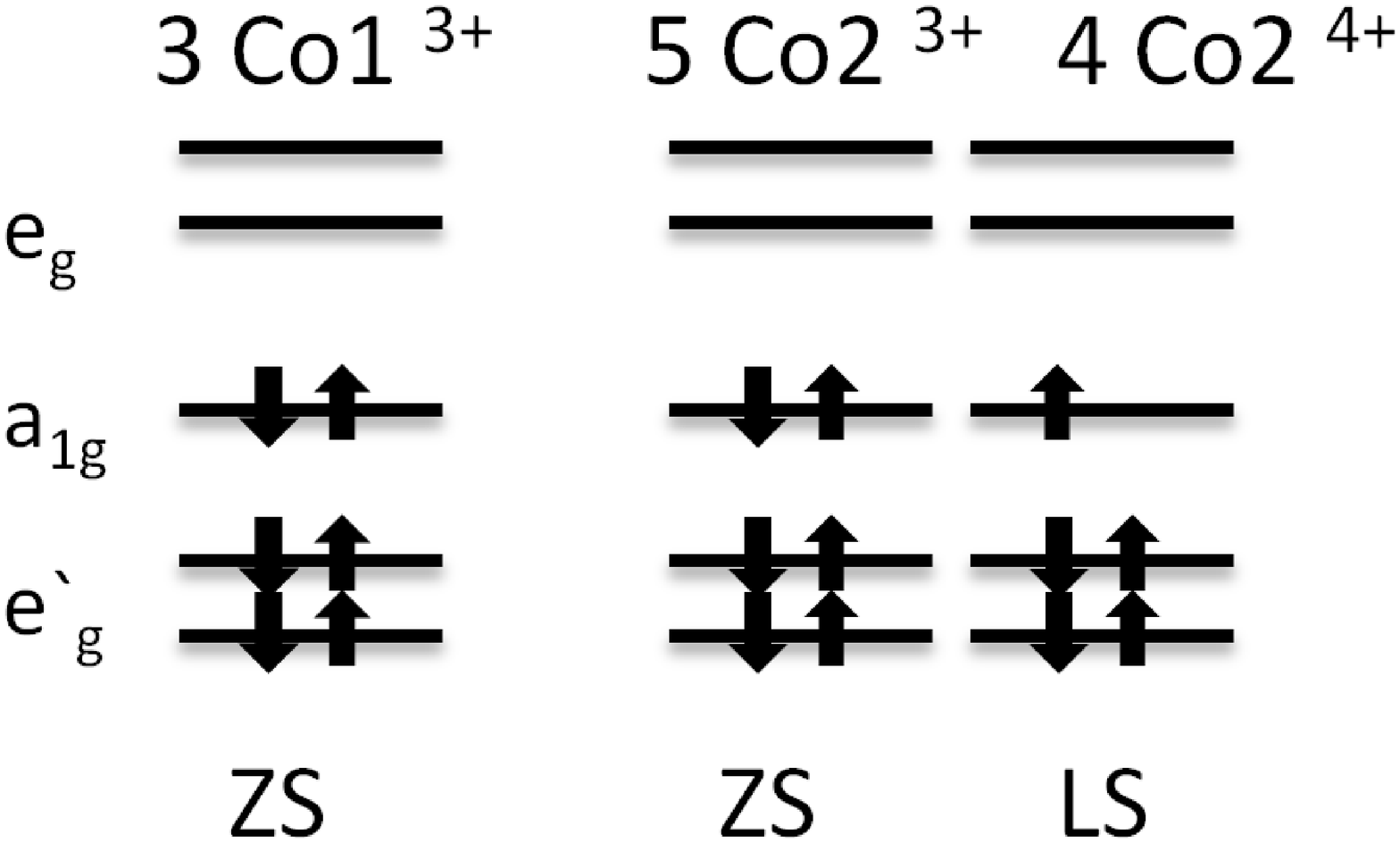}}
\end{minipage}
\hfill
\begin{minipage}[h]{0.5\linewidth}
\textit{(b)}
\center{\includegraphics[height=100px]{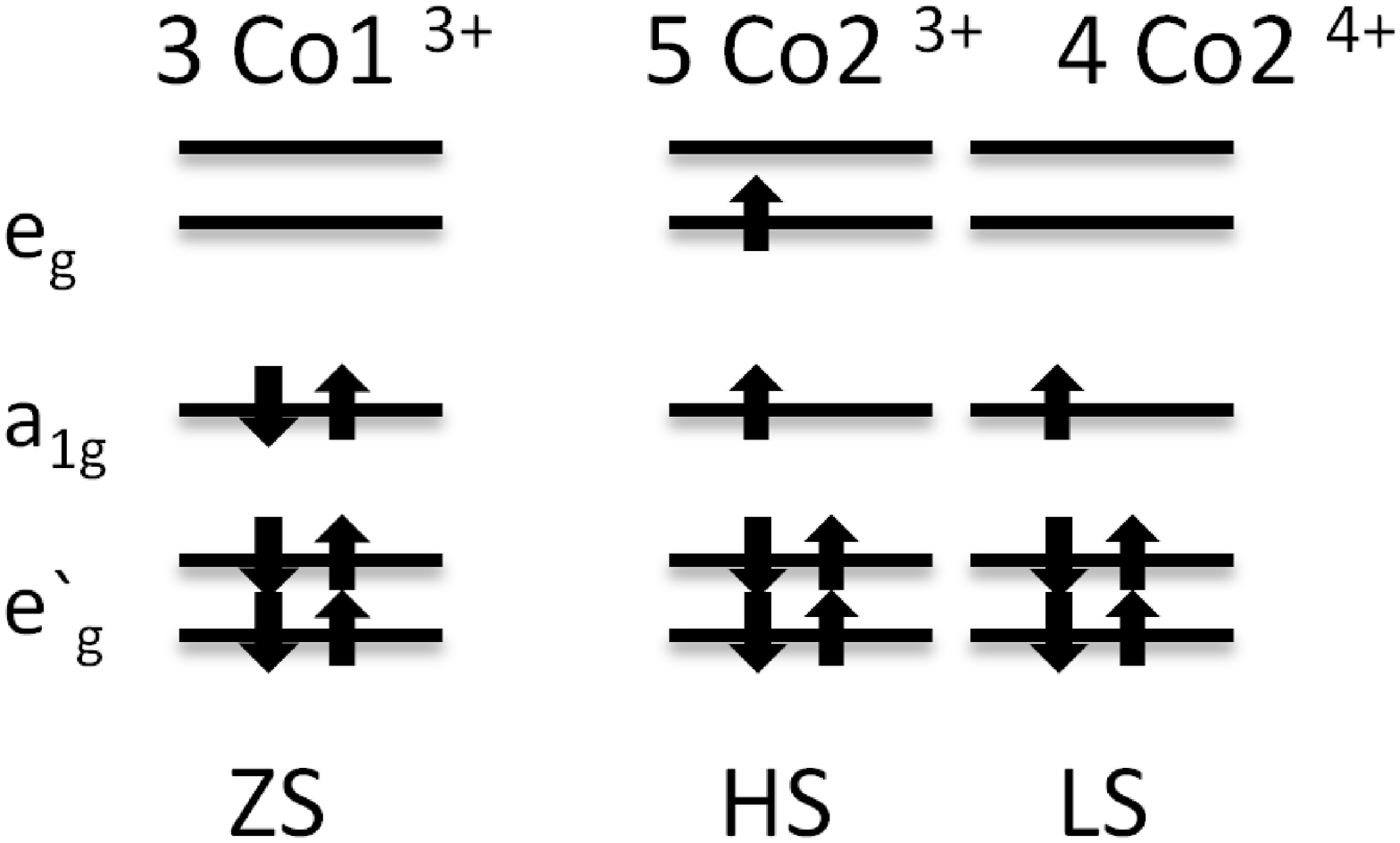}}
\end{minipage}
 \caption{Electronic configuration diagram of one symmetry irreducible layer of Na${}_{2/3}$CoO${}_2$ in low spin\textit{(a)} and intermediate spin \textit{(b)} states. The total magnetic moment of layer is 4 $\mu_\textrm{B}$ and 14 $\mu_\textrm{B}$, correspondingly. }
 \label{fig:Na23CoO2ElectronDiagram}
\end{figure}

Schematic electronic configurations for one symmetry irreducible layer of Na${}_{2/3}$CoO${}_2$ in low spin and intermediate spin states are depicted on Figure~\ref{fig:Na23CoO2ElectronDiagram}.

 We perform geometric optimization of elementary cell in both the LS and IS spin states within GGA and GGA+U methods, within range of $U$ from 1 to 6 eV. It should be noticed, that even without symmetrization of charge densities and forces, the arrangement of sodium ions remains the same. Thus, the stability of proposed sodium spatial distribution is confirmed. The energies of system under consideration in FM and AFM interlayer order in LS state have been calculated. The energies comparison indicates that the AFM order has less energy of about 7.5 meV per one irreducible layer and thus is more favorable.

The charge/spin separation for optimized structures in LS and IS states was compared. We start from the case of the absence of strong correlation between 3d electrons (the Hubbard parameter is equal to $U=0$). Then the GGA provides a uniform with FM spin-polarized metallic LS state of Co-planes. In turn, increasing of Hubbard repulsion $U$ parameter results in the localization of magnetic moments at Co2 sublattice when $U$ exceeds $\approx 3$ eV. The spin state is obtained to be non-uniform: in the LS state one finds the robust magnetic moment of about 0.6 $\mu_{\textrm{B}}$ at the Co2a(b) sites, forming the kagome superlattice, while Co1 sites remain almost spinless. Thus a kind of a "spin separation" between the sites in the triangular Co lattice is appeared. Such a type of separation resembles the spin separation indicated by the spin-polarized GGA+U calculations for Na${}_{0.5}$CoO${}_2$~\cite{Li2005a}.

For the IS state the situation is more complicated. It is not possible to stabilize IS state for the low values of Hubbard $U$ parameter. The minimal value of Hubbard parameter at which the IS state was obtained  is $U$ = 4 eV. This value is close to the $U = 5$ eV reported in \cite{Shorikov2011}, which is necessary to stabilize high spin state in Na${}_{0.61}$CoO${}_2$. So, one can maintain that our calculation is coherent with the other ab-initio calculations.

\begin{figure}[h]
\begin{minipage}[h]{0.47\linewidth}
\textit{(a)}
 \center{
 \includegraphics[width=1 \linewidth]{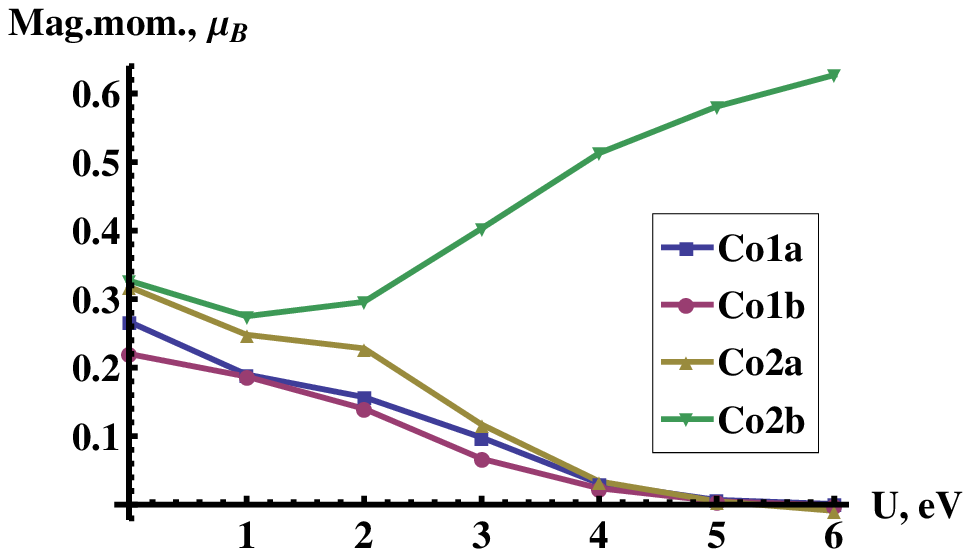}}
 \end{minipage}
\hfill
\begin{minipage}[h]{0.47\linewidth}
  \textit{(b)}
 \center{  \includegraphics[width=1\linewidth]{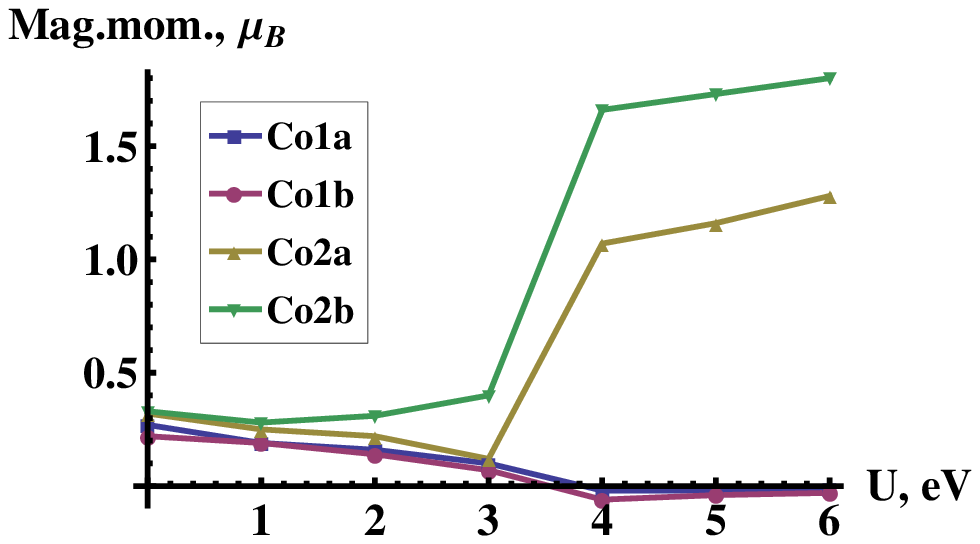}}
\end{minipage}
  \caption{Magnetic moments of cobalt ions depends on on-site Hubbard U correction parameter in low \textit{(a)} and intermediate \textit{(b)} spin states.}\label{fig:UVar-MM-LSIS}
\end{figure}

Spin separation curves for IS state are depicted on Figure~\ref{fig:UVar-MM-LSIS},\textit{(b)}. One can conclude that the increasing of Hubbard $U$ parameter leads to the growth of magnetic moment at cobalt ions in "favorable" crystallographic positions (Co2b and/or Co2a), i.e. to the localizations of holes in Co2 sublattice.

\begin{figure}[h!]
\begin{minipage}[h]{0.47\linewidth}
\textit{(a)}
 \center{
 \includegraphics[width=1.00\linewidth]{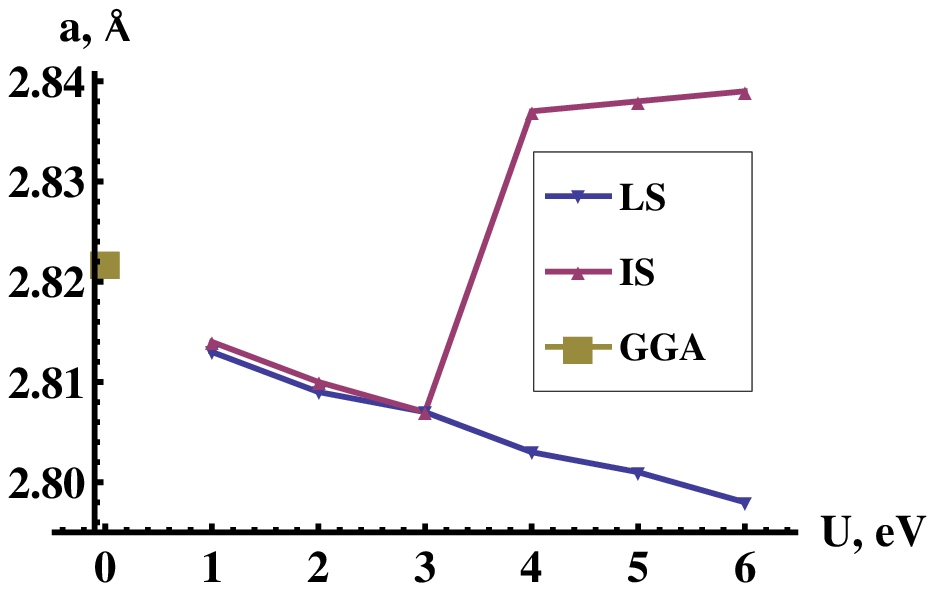}}
 \end{minipage}
\hfill
\begin{minipage}[h]{0.47\linewidth}
  \textit{(b)}
 \center{
  \includegraphics[width=1.00\linewidth]{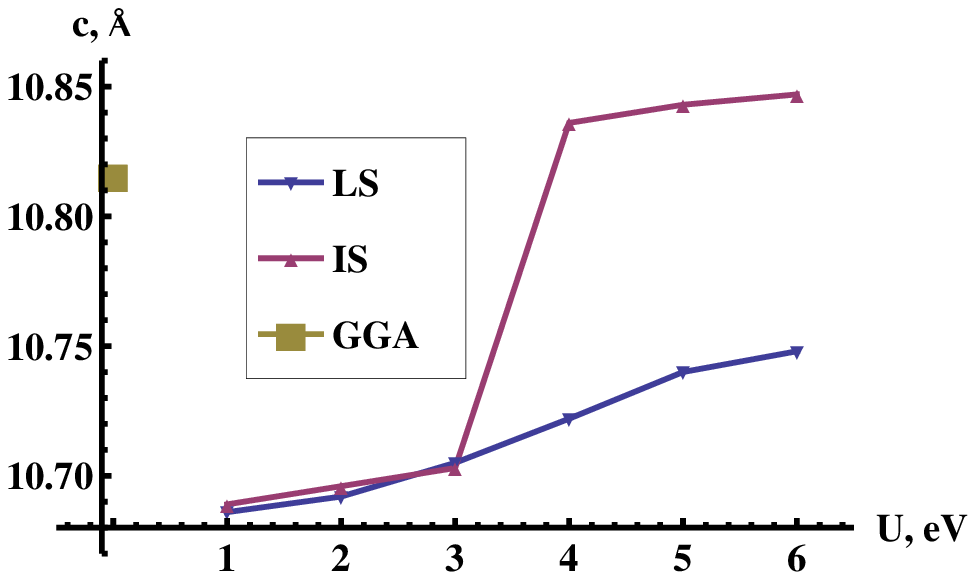}}
\end{minipage}
 \caption{Lattice parameters $a$ \textit{(a)} and $c$ \textit{(b)} in a dependence on on-site Hubbard U correction parameter in LS and IS states. "GGA" denotes the pure GGA method, without LSDA+U correction for LS state.}\label{fig:UVar-AC-LSIS}
\end{figure}

For the geometric parameters, such as the lattice constants $a$ and $c$, there is no single motif. Increasing of $U$ leads to decreasing of hexagonal inplane lattice parameter $a$ in LS state, but small growth in IS state (Figure~\ref{fig:UVar-AC-LSIS}, \textit{(a)}), whereas in the direction which is perpendicular to CoO${}_2$ planes the lattice constant $c$ demonstrates monotonic growth for both spin states (Figure~\ref{fig:UVar-AC-LSIS}, \textit{(b)}). The maximal relative difference of lattice constants between two spin states is 1.5\% for $a$ parameter and 0.9 \% for $c$ parameter, which lies in the usual DFT error range of 2-3 \%. Probably, one can explain such behavior by redistribution of electron charge density from inplane toward interplane orientation, which leads to repulsion of layers. Significant increasing of both lattice parameters in IS state can be explained in the following way: in the IS state electrons on cobalt ions occupy the upper $e_g$ bands. These bands are geometrically oriented toward nearest oxygen ions, which has partial negative charge and, as results, tends to repulse from cobalt ions, which leads to increasing of both $a$ and $c$ lattice parameters.

Experimental lattice parameters for sodium concentration $x$ close to $2/3$ are presented in the Table~\ref{table:Na067CoO2-geometry}. As it can be seen, experimental lattice parameters are closer to IS state, but the relative difference of order 2 \% lies in usual range of geometrical parameters error, calculated within DFT methods. Thus, the lattice parameter can not be a strict criteria to identification of experimental spin state.
\begin{table}
  \caption{\label{table:Na067CoO2-geometry}Experimental lattice parameters for Na${}_{x}$CoO${}_2$ elementary cell.}
\begin{tabular}{@{}llll}
  \hline
   Compound & Ref. &a(\AA) & c(\AA)\\
  \hline
  Na${}_{0.61}$CoO${}_2$ & \cite{Jorgensen2003} & 2.832 & 10.843 \\
  Na${}_{0.66}$CoO${}_2$ & \cite{Kroll2006} & 2.837 & 10.97 \\
  Na${}_{0.67}$CoO${}_2$ & \cite{mukhamedshin2007influence} & - & 10.938 \\
  \hline
\end{tabular}
\end{table}

\begin{figure}
\centering  \includegraphics[width=0.55 \linewidth]{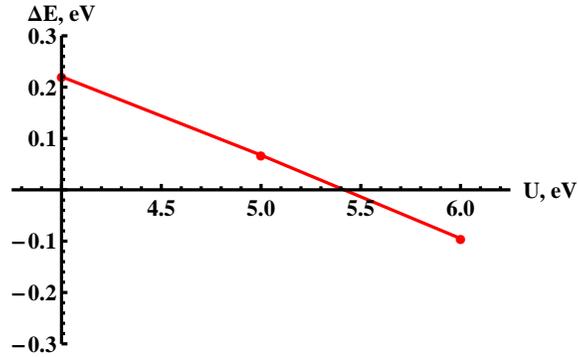}\\
  \caption{Difference between the ground-state energy of the system in IS and LS states in a dependence on the parameter of on-site Hubbard interaction parameter $U$ . Here, the energy is calculated per one cobalt ion in the Co2 crystallographic position. When $U$ is less than 4 eV, the IS state is unstable and only the LS state is accessible.}\label{fig:UVar-deltaE}
\end{figure}

The natural question is rising - which spin configuration is more favorable - the low spin state or intermediate spin state? The energy difference (per Co2 ion) between these two states depended on on-site Hubbard correction parameter $U$ is depicted on the Figure~\ref{fig:UVar-deltaE}. In the range 0 eV $< U <$ 3 eV it is not possible to stabilize IS state (simply, it does not exist), which shown as zero energy difference. For the values of $U$ equal to 4 and 5 eV LS spin state has less energy, whereas for higher $U$ value the situation is inverse. But neither zero point energy nor finite temperature effect was not taken into account, whereas, in fact, they can influence onto the ground state determination.

\begin{figure}
\begin{minipage}[h]{0.49\linewidth}
\textit{(a)}
\center{
\includegraphics[width=\linewidth]{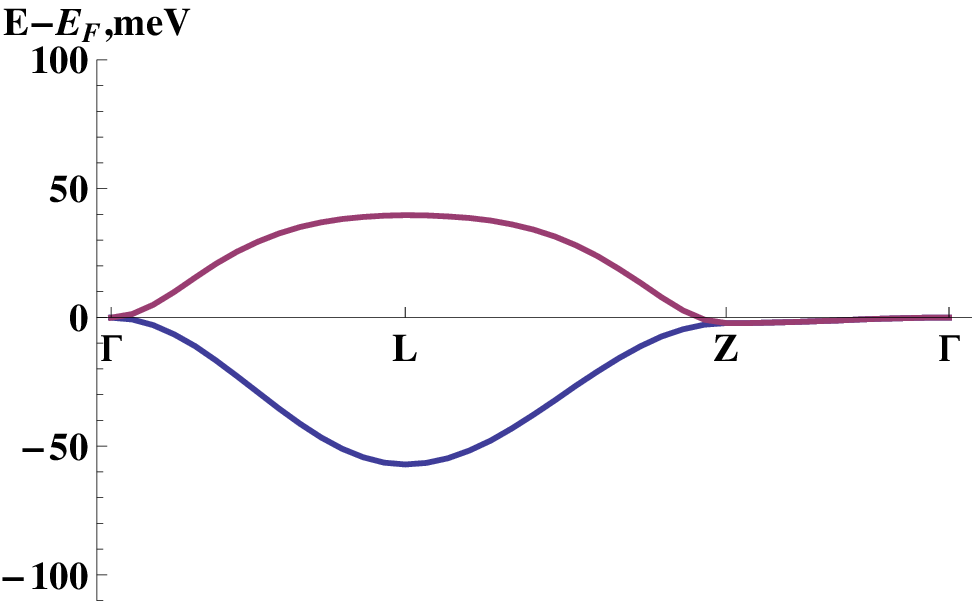}}
\end{minipage}
\hfill
\begin{minipage}[h]{0.49\linewidth}
\textit{(b)}
\center{\includegraphics[width=\linewidth]{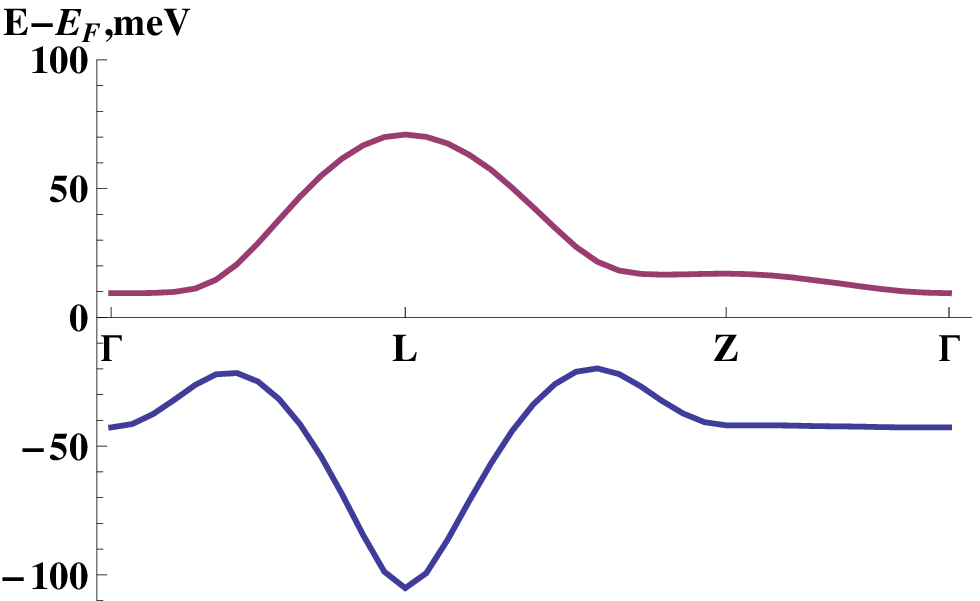}}
\end{minipage}
  \caption{Band structure for LS \textit{(a)} and IS \textit{(b)} states about the Fermi level. The LS state demonstrates the metallic character, while IS shows a small insulating band gap ($\approx$ 30 meV). Here, the on-site Hubbard interaction parameter $U$ = 4 eV.}\label{fig:Na066-LSIS-Band}
\end{figure}

The band structures calculated for both of the spin states are presented on Figure~\ref{fig:Na066-LSIS-Band}. IS state possess a small band gap about 30 meV at the Fermi level. Whereas bands structure of the LS state has the definitive metallic character for all the range of $U$. Experimentally, the metallic state of the cobaltate was confirmed by the measurement of the resistivity~\cite{Hasan2004}.

\begin{figure}
\centering
\includegraphics[width=0.45 \linewidth]{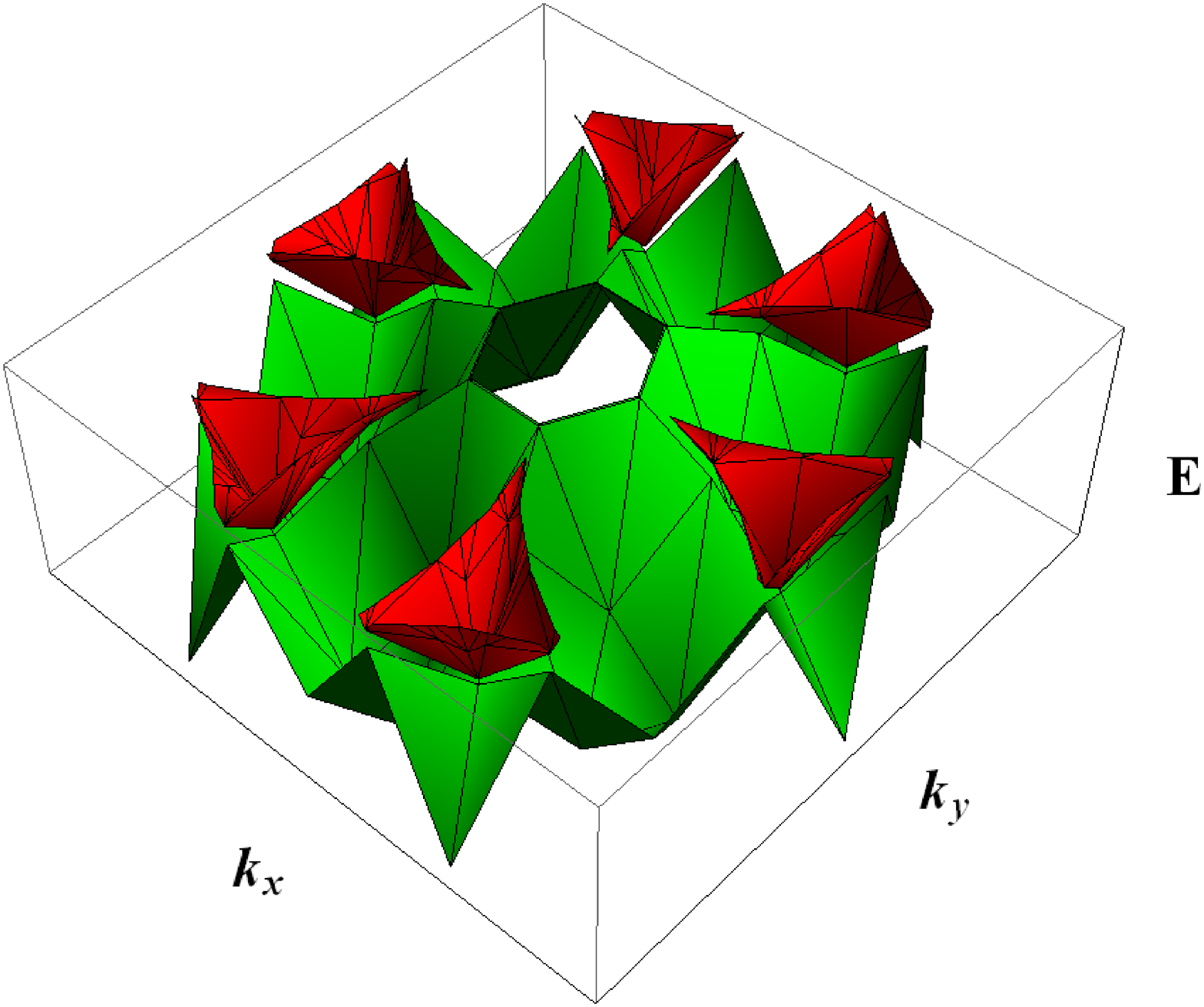}
\hfill
\includegraphics[width=0.45 \linewidth]{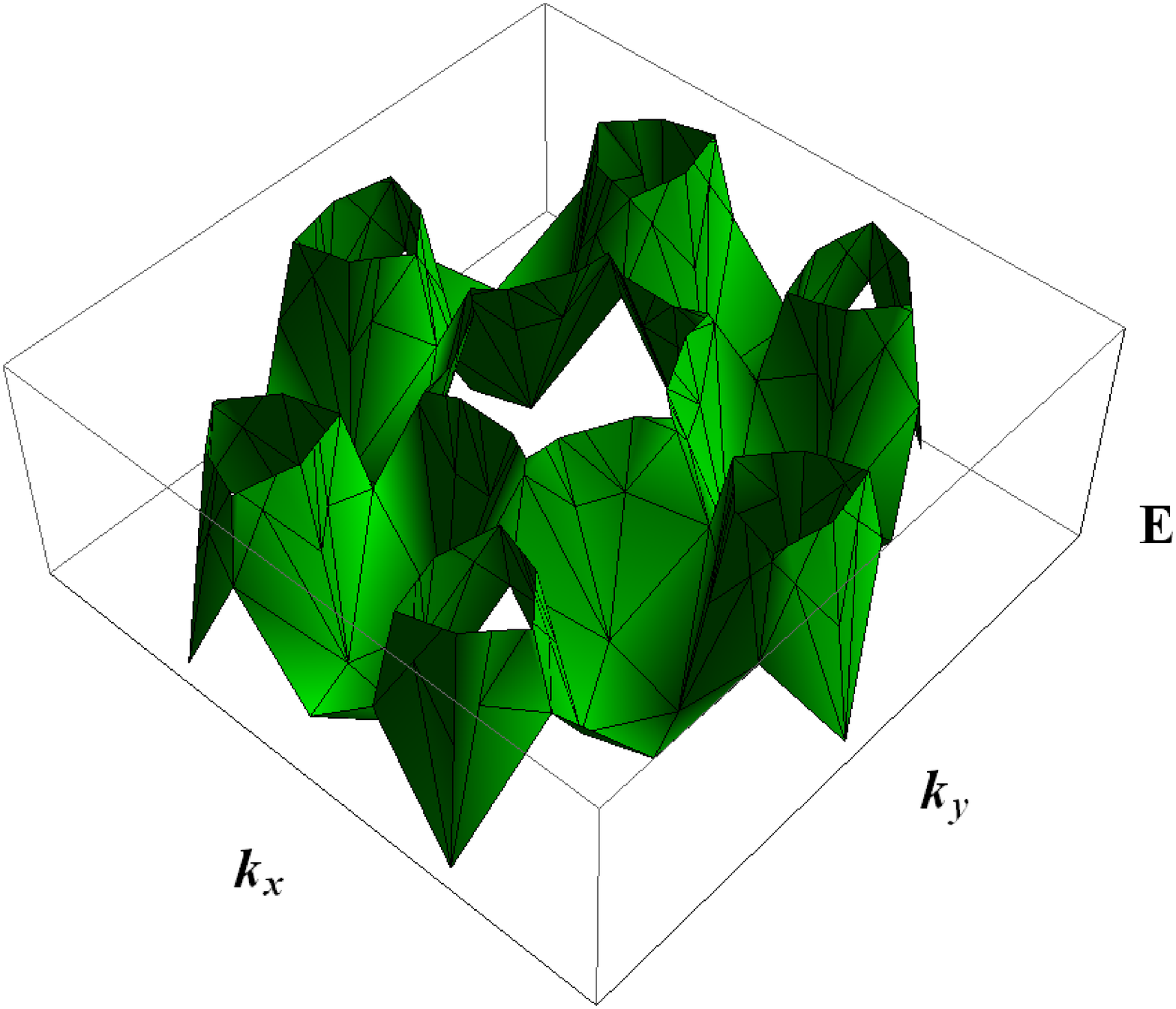}\\

\includegraphics[width=0.40 \linewidth]{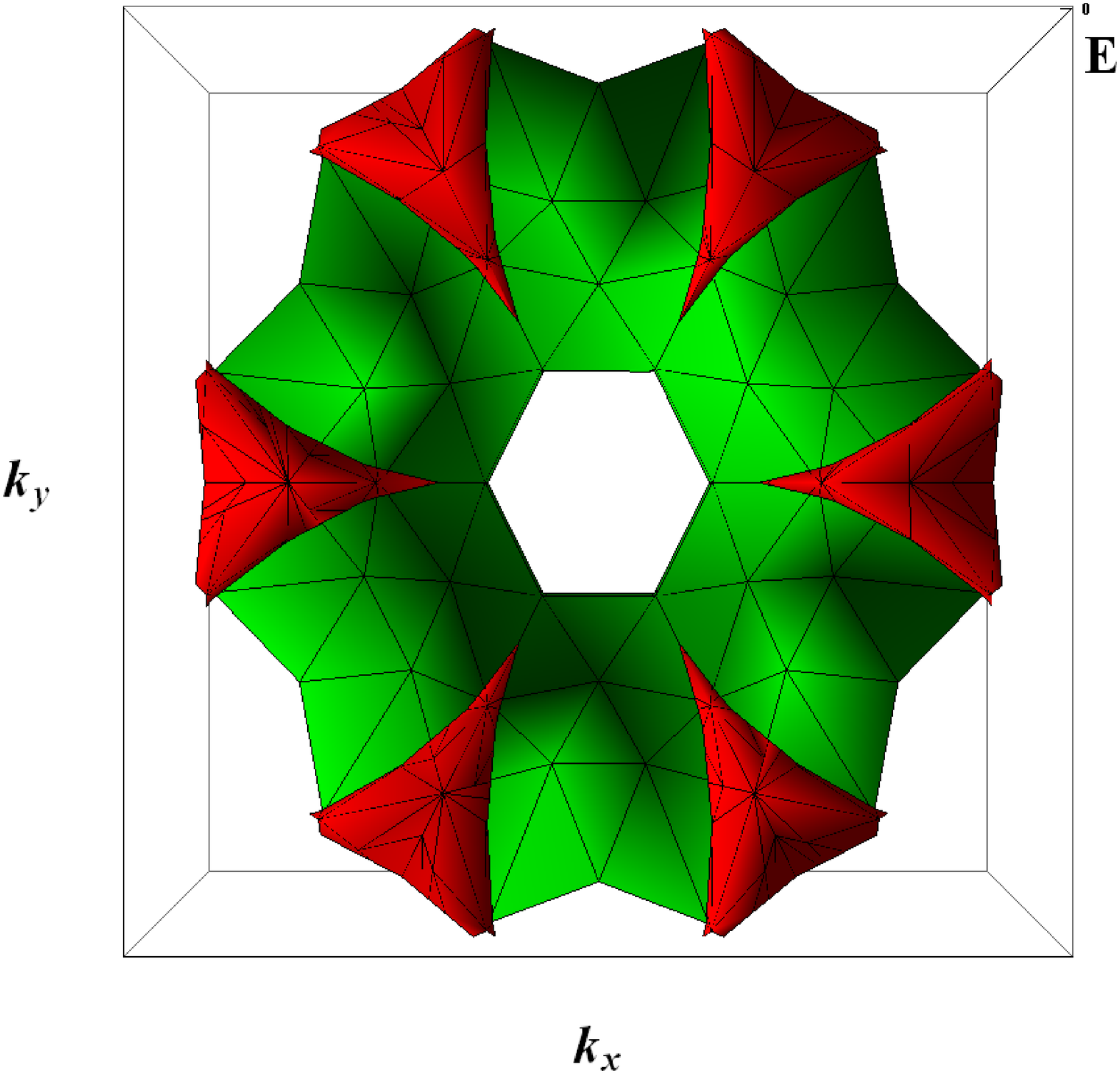}
\hfill
\includegraphics[width=0.40 \linewidth]{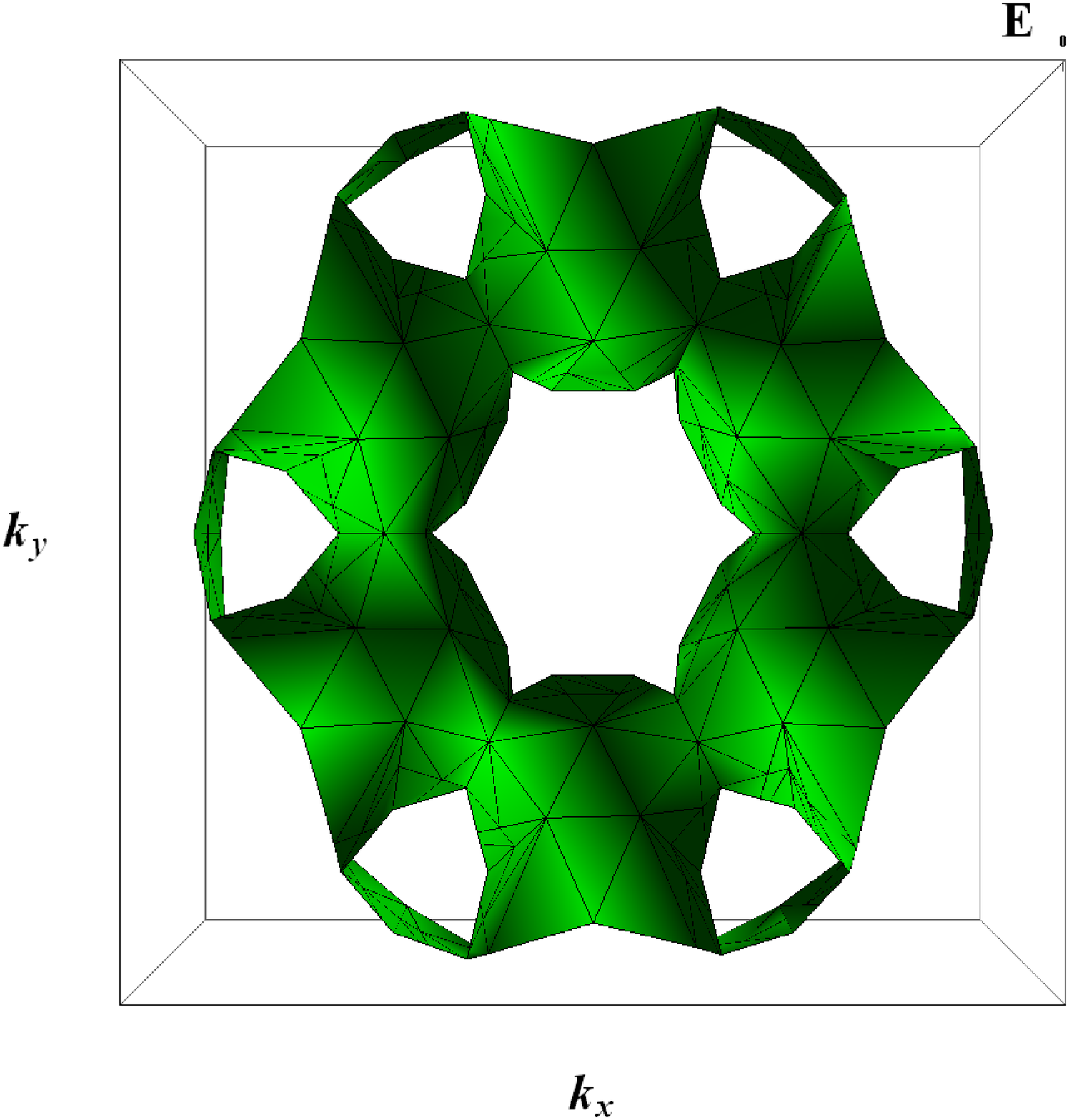}\\
  \caption{Band structure near Fermi level for LS state in $XY$ plane for two cases: $U$=1 eV (left) and $U$=6 eV(right). On the left figure, Fermi level is crossed by two bands (green and red), while on the right, only by one band (green).  Moreover the six small pockets change its character from the electron-like (concave) to the hole-like (convex).}\label{fig:Na066-LSIS-Fermi}
\end{figure}

To calculate Fermi surface (FS) the Brillouin zone (BZ) was sampled with $9 \times 9 \times 9$ k-mesh. Since band structures show no variation of energy in $z$ direction, perpendicular to CoO${}_2$ plane ($Z-G$ line on the Figure~\ref{fig:Na066-LSIS-Band} is straight), we are interested in distribution of energy in $XY$ plane. On Figure~\ref{fig:Na066-LSIS-Fermi} band structures of LS state, cut at Fermi level for two different values of $U$ ($U$=1 eV and $U$ = 6 eV) depends on $k_x$ and $k_y$ are depicted. Because of the relatively rough k-spacing in Brilloiun zone, some numerical interpolation artefacts are arised. Nevertheless, one can note the tendency of increasing of volume of central Fermi pocket and decreasing of volume of additional small Fermi pockets, in accordance with the Luttinger theorem. But the six small pockets in Fermi surface remains independently to the value of the Hubbard $U$ parameter.

Moreover, while central pocket on Fermi surface always has hole character(convex), the small side pockets change their natures from electron-like(concave) at $U=1$ eV to hole-like(convex) at $U=6$ eV, as depicted on Figure~\ref{fig:Na066-LSIS-Fermi}.
For $U$=1 eV, electron velocity at the central FS pocket is about $v_{\mathrm{F}} \sim 4\times 10^4$ m/s, whereas for sodium concentration $x=1/3$ $v_{\mathrm{F}} \sim 6 \times 10^4$ m/s. The size of FS central pocket in our case is about $k_{\mathrm{F}} \sim 0.1$ \AA${}^{-1}$. This value is less compared to $x \sim 1/3$ case (where $k_{\mathrm{F}} \sim 0.7$ \AA${}^{-1}$ \cite{Qian2006}), but it should be noted, that the elementary cell at $x=2/3$ is larger, which should results in decreasing of first BZ size, and thus the size of FS, deduced from the experiment.

Analysis of macroscopic magnetic susceptibility of Na${}_{2/3}$CoO${}_2$ sample in temperature range from 50 K to 250 K in Ref.~\cite{mukhamedshin2007influence} leads to effective magnetic moment $\mu_\mathrm{eff}=1.12 \mu_B$ per Co2 site, whereas effective magnetic moment for LS state is only about $ 0.3 \mu_B$ per Co2 site. One could suggest that the above-mentioned experimental data should correspond to IS state. But, firstly, these data indicate only the average magnetic moment and extra magnetic moments can be induced by some impurities of sample. Secondly, IS state has a small band gap and should not demonstrate metallic conductivity (linear dependence of resistance on temperature) as it observed for Na${}_{0.7}$CoO${}_2$ \cite{Hasan2004} for wide temperature range.

\section{Discussion}

The existence of the strong FM correlations for the Hubbard model on the trigonal lattice while hole concentration corresponds to $x =2/3$  has been shown in Ref.~\cite{Boehnke2010}. Such a system possess the instability with respect to simultaneous appearance of charge and spin density waves with periodicity of Kagome superlattice. The probable reason of a such type of instability is in van-Hoove singularity of the density of states of $3d$-electrons nearby the Fermi level on the Kagome lattice~\cite{Peil2011}. This explains why the extensional wave function of the Co${}^{4+}$ hole with spin $s=1/2$ tends to localize from the whole layer into the sublattice of Co2: it leads to decreasing of the magnetic energy due to the FM transition and thus, optimization of on-site Hubbard (Coulomb) correlations. At the same time, the inert states Co${}^{3+}$ (s=0) are superseded to the residuary Co1 sites. In turn, it makes the next to the Co1 ions Na1 sites attractive for the Na${}^{+}$ ions. Then the Coulomb interaction between the sodium sublattice and cobalt sublattice tends to pin the whole charge order in the system. Such a scheme could be the probable scenario of stabilization in a metallic system: FM of a ground state with concurrent charge ordering of Co${}^{3+}$ (Co${}^{4+}$) ions on the Co1 (Co2) sites and arrangement of the Na${}^{+}$ ions on the Na1 and Na2 sites.

Charge and spin separation for cobalt ions has been demonstrated in the present work. An AFM-A type magnetic order (antiferromagnetic interplane and ferromagnetic inplane) has been derived as more energetically stable. We have studied two possible spin states - low spin state and intermediate spin state. Obtained IS state resemble a spin-polaron excitation state, proposed by Khaliullin \textit{et al.} ~\cite{Khaliullin2008a,chaloupka2008}, but this state is an insulator state with a small band gap of about 30 meV. Our calculations have indicated that for small value of the Hubbard parameter $U < 6$ eV the ground state is a LS state, whereas at $U = 6$ eV IS state has less energy. Also, the charge/spin separation dependence on the Hubbard $U$ parameter has been observed.

The band theory predict the complicated Fermi surface (FS) with one large hole pocket around the $\Gamma = (0, 0, 0)$  point and six small pockets near the $K =(0,\frac{4 \pi}{3},0)$ points of the hexagonal Brillouin zone at least for $x < 0.5$ \cite{Singh2000}. However, intensive investigations by several ARPES groups reveal absence of six small pockets in both \NaxCoO *$y$H${}_2$O and in its parent compound \NaxCoO \cite{Yang2005,Zhang2004,Hasan2004}.

Independently on the value of on site  Hubbard correlation energy $U$, the calculated Fermi surface has six small pockets that are not presented in the ARPES experiment~\cite{Hasan2004}. The disagreement between ARPES spectra and ab initio calculated band structure points to the importance of the electronic correlations in these oxides. Possible cause of this unconformity is that the ARPES experiment shows electronic states of the sample surface, whereas, in certain cases, they can differ significantly from the bulk states \cite{Pillay2008a}. Indeed, independent experiments on transport properties which exploits  Shubnikov - de Haas effect in \NaxCoO for $x=0.71$ and $0.84$, exhibit the presence of the Fermi surface pockets \cite{Balicas2008}.
Thus, for the correct understanding of contradiction between ARPES experiment and ab initio calculations of realistic crystal cell, an ab initio studying of sodium cobaltate surfaces should be done.

Comparison of calculated results with the experimental data on conductivity and Fermi surface indicates that ground state is a LS state. At the same time, the macroscopic magnetic susceptibility measurements indicate the high magnetic moment per cobalt ion and thus one can not conclude that one has the IS of sample.  Nevertheless, it is more probable, that the extra magnetic moments are not appeared from Na${}_{2/3}$CoO${}_2$ compound, but they can be induced by some impurities of the sample or defects of crystal lattice (such as sodium vacancies \cite{Alloul2009a}).

\section{Conclusion}

In the present work we have modeled electronic and magnetic properties of sodium cobaltate Na${}_{2/3}$CoO${}_2$ within GGA+U approximation. In contrast to previous work~\cite{Zhang2004,Zhou2005,Korshunov2007}, for the first time, a realistic, experimentally obtained elementary cell of this compound~\cite{Alloul2009a} was used. Ab initio calculations indicate that proposed sodium imprint is stable in the sense that it does not change after the full structure relaxation.

The microscopic theory of  FM metallic state of Na${}_{2/3}$CoO${}_2$ with a real Na${}^{+}$ arrangement in Na1 and Na2 sites~\cite{Alloul2009a} was developed. It was shown that at $U>3$eV in the system the complex ordering, which combines the redistribution of Co${}^{4+}$ into Co2 sublattice, arises. The calculated spatial structure of the electronic ground state is coherent to the experimentally deduced~\cite{Alloul2009a}. It was established that the presence of both electrostatic potential induced by the periodic Na${}^{+}$ arrangement and the strong Coulomb correlations in the 3d orbitals is sufficient to appearance of such ground state. Thus, the tricky ground state of the cobalte could be the result of the peculiar cooperation between two relevant interacting subsystems: the patterning ions and the strongly correlated 3d electrons of the Co-layers.

\section*{Acknowledgments}
Authors would like to thank Prof. Irek Mukhamedshin for his NMR/NQR experimental data and model of crystal structure. Also, Y.L. would like to thank Region Pays de la Loire for the partial financial support (MOA119). A support by the RFBR under the project No. 10-02-01005-a is acknowledged as well.

\bibliography{cobaltate}

\providecommand{\newblock}{}
\begin{thebibliography}{10}
\expandafter\ifx\csname url\endcsname\relax
  \def\url#1{{\tt #1}}\fi
\expandafter\ifx\csname urlprefix\endcsname\relax\def\urlprefix{URL }\fi
\providecommand{\eprint}[2][]{\url{#2}}

\bibitem{Fujita2001}
Fujita K, Mochida T and Nakamura K 2001 {\em Jpn. J. Appl. Phys.\/} {\bf 40}
  4644

\bibitem{Terasaki1997}
Terasaki I, Sasago Y and Uchinokura K 1997 {\em Phys. Rev. B\/} {\bf 56}
  R12685--R12687

\bibitem{takada2003superconductivity}
Takada K, Sakurai H, Takayama-Muromachi E, Izumi F, Dilanian R and Sasaki T
  2003 {\em Nature\/} {\bf 422} 53--55

\bibitem{Marianetti2007a}
Marianetti C and Kotliar G 2007 {\em Phys. Rev. Lett.\/} {\bf 98} 176405

\bibitem{Lang2008}
Lang G, Bobroff J, Alloul H, Collin G and Blanchard N 2008 {\em Phys. Rev. B\/}
  {\bf 78} 155116

\bibitem{Meng2008a}
Meng Y, Hinuma Y and Ceder G 2008 {\em J. of chem. phys.\/} {\bf 128} 104708
  ISSN 0021-9606

\bibitem{Hinuma2008}
Hinuma Y, Meng Y and Ceder G 2008 {\em Phys. Rev. B\/} {\bf 77} 224111

\bibitem{Meng2005}
Meng Y, {Van der Ven} a, Chan M and Ceder G 2005 {\em Phys. Rev. B\/} {\bf 72}
  172103 ISSN 1098-0121

\bibitem{Singh2000}
Singh D~J 2000 {\em Phys. Rev. B\/} {\bf 61} 13397--13402

\bibitem{Alloul2009a}
Alloul H, Mukhamedshin I~R, Platova T~A and Dooglav A~V 2009 {\em EPL\/} {\bf
  85} 47006

\bibitem{Zhou2005}
Zhou S, Gao M, Ding H, Lee P~A and Wang Z 2005 {\em Phys. Rev. Lett.\/} {\bf
  94} 206401

\bibitem{Zhang2004}
Zhang P, Luo W, Cohen M and Louie S 2004 {\em Phys. Rev. Lett.\/} {\bf 93}
  236402

\bibitem{Blochl94}
Blochl P 1994 {\em Phys. Rev. B\/} {\bf 50} 17953

\bibitem{Kresse96}
Kresse G and Furthmuller J 1996 {\em Phys. Rev. B\/} {\bf 54} 169

\bibitem{PBEsol}
Perdew J, Ruzsinszky A, Csonka G and Vydrov O 2008 {\em Phys. Rev. Lett.\/}
  {\bf 100} 136406

\bibitem{Dudarev98}
Dudarev S, Botton G, Savrasov S, Humphreys C and Sutton A 1998 {\em Phys. Rev.
  B\/} {\bf 57} 1505

\bibitem{Zhou2004}
Zhou F, Cococcioni M, Marianetti C~A, Morgan D and Ceder G 2004 {\em Phys. Rev.
  B\/} {\bf 70} 235121

\bibitem{Zhou.PRB.69.201101}
Zhou F, Marianetti C~A, Cococcioni M, Morgan D and Ceder G 2004 {\em Phys. Rev.
  B\/} {\bf 69}(20) 201101

\bibitem{Li2005a}
Li Z, Yang J, Hou J~G and Zhu Q 2005 {\em Phys. Rev. B\/} {\bf 71} 024502

\bibitem{Shorikov2011}
Shorikov a, Korshunov M~M and Anisimov V~I 2011 {\em JETP Letters\/} {\bf 93}
  80--85

\bibitem{Jorgensen2003}
Jorgensen J~D, Avdeev M, Hinks D~G, Burley J~C and Short S 2003 {\em Phys. Rev.
  B\/} {\bf 68} 214517

\bibitem{Kroll2006}
Kroll T 2006 {\em {On the electronic structure of layered sodium cobalt
  oxides}\/} Ph.D. thesis

\bibitem{mukhamedshin2007influence}
Mukhamedshin I, Alloul H, Collin G and Blanchard N 2007 {\em Arxiv preprint
  cond-mat/0703561\/}

\bibitem{Hasan2004}
Hasan M~Z, Chuang Y~D, Qian D, Li Y~W, Kong Y, Kuprin A, Fedorov A~V,
  Kimmerling R, Rotenberg E, Rossnagel K, Hussain Z, Koh H, Rogado N~S, Foo M~L
  and Cava R~J 2004 {\em Phys. Rev. Lett.\/} {\bf 92} 246402

\bibitem{Qian2006}
Qian D, Wray L, Hsieh D, Viciu L, Cava R~J, Luo J, Wu D, Wang N and Hasan M
  2006 {\em Phys. Rev. Lett.\/} {\bf 97} 186405 ISSN 0031-9007

\bibitem{Boehnke2010}
Boehnke L and Lechermann F 2010  (\textit{Preprint} \eprint{1012.5943})

\bibitem{Peil2011}
Peil O, Georges A and Lechermann F 2011 {\em Phys. Rev. Lett.\/} {\bf 107}
  236404

\bibitem{Khaliullin2008a}
Khaliullin G and Chaloupka J 2008 {\em Physical Review B\/} {\bf 77} 104532

\bibitem{chaloupka2008}
Chaloupka J and Khaliullin G 2008 {\em Arxiv preprint arXiv:0806.1682\/}

\bibitem{Yang2005}
Yang H, Pan Z, Sekharan A, Sato T, Souma S, Takahashi T, Jin R, Sales B,
  Mandrus D, Fedorov A, Wang Z and Ding H 2005 {\em Phys. Rev. Lett.\/} {\bf
  95} 146401

\bibitem{Pillay2008a}
Pillay D, Johannes M and Mazin I 2008 {\em Phys. Rev. Lett.\/} {\bf 101} 246808

\bibitem{Balicas2008}
Balicas L, Jo Y, Shu G, Chou F and Lee P 2008 {\em Phys. Rev. Lett.\/} {\bf
  100} 126405 ISSN 0031-9007

\bibitem{Korshunov2007}
Korshunov M, Eremin I, Shorikov A, Anisimov V~I, Renner M and Brenig W 2007
  {\em Phys. Rev. B\/} {\bf 75} 094511

\end{thebibliography}

\end{document}